\documentclass{aa}
\usepackage{graphicx}
\usepackage{natbib}
\usepackage{color}
\bibpunct{(}{)}{;}{a}{}{,} 
\begin{document}

\def\ms{M$_{\odot}$}
\def\zs{Z$_{\odot}$}
\def\mi{M$_{\rm IN}$}
\def\zi{Z}
\def\xi{Z$_{\rm IN}$}
\def\me{M$_{\rm ENV}$}
\def\nsn{N$_{\rm Ib,c}$/N$_{\rm II}$}
\def\mup{M$_{\rm Ib,c}$}
\def\lb{L$_{\rm B,\odot}$}
\def\be{$\beta^+$}
\def\ee{e$^+$}
\def\ne{N$_{e^+}$}
\def\co{$^{56}$Co }
\def\ti{$^{44}$Ti }

\def\ls{$^{6}$Li }
\def\lv{$^{7}$Li }
\title{The energetics, evolution, and stellar depletion of $^6$Li in the early Galaxy}

\author{ Nikos Prantzos
        }

\authorrunning{N. Prantzos}
 
\titlerunning{$^6$Li in the Galaxy}


\institute{ Institut d'Astrophysique de Paris, 98bis Bd. Arago, 
              75104 Paris, France,
                \email{prantzos@iap.fr}
           }
\date{accepted: October 7, 2005}

\abstract{Motivated by the recent report of a high $^6$Li ``plateau''
extending to low metallicities in Galactic halo stars, we study
the energetics of an early production of \ls through the interaction
of energetic particles  with the interstellar medium. We then explore 
the potential of various candidate
sources of pre-galactic energetic particles and show that, in general, they  
fail to satisfy the observational and theoretical requirements
(especially if the \ls \ plateau is at a considerably higher 
level than observed). Succesful candidates appear to be:
supernova explosions with abnormally low metal yield;  the 
massive black hole in the Galactic center (provided that it was
formed early on and that it was then radiating much more efficiently 
than today); and, perhaps,  an early accretion phase of
supermassive black holes in galaxies.
Assuming that \ls \ is indeed pre-galactic, 
we study  the galactic
evolution of the light isotopes $^7$Li, \ls, and Be in a self-consistent way.
We find that the existence of a \ls \ plateau is hard to justify, 
unless a fine-tuned  and metallicity-dependent depletion mechanism 
of \ls \ in stellar envelopes is invoked. The depletion of \ls should be
different, both in magnitude and in its metallicity dependence, than 
the depletion required to explain current observations of Li 
(mostly $^7$Li) in halo stars.  
If the recently reported \ls ``plateau'' is confirmed, 
our analysis suggests important implications for 
our understanding of the production, evolution, and stellar depletion
of the Li isotopes.
}


\maketitle
%

\section{Introduction}

The idea that the light and fragile elements Li, Be and B are
produced by the interaction  of the energetic nuclei
of galactic cosmic rays (CRs) with the nuclei of the interstellar medium
(ISM) was introduced 35 years ago (Reeves et al. 1970, Meneguzzi et al. 1971,
hereafter MAR).
In those early works it was shown that, by taking into account
the relevant spallation cross-sections 
and with plausible assumptions about the
CR properties (injected and propagated spectra, intensities etc.; see
Sect. 2) one can reproduce the abundances of those light
elements observed in meteorites  and  in CRs, i.e after $\sim$10$^{10}$ yr
of galactic evolution (see, e.g. Reeves 1994, for a review on LiBeB). 
However, the earliest evolution of the light
element abundances was not considered in those works.

One of the major cosmological developments of the 1980s was the
discovery  of the Li plateau in low metallicity halo stars
(Spite and Spite 1982).
The unique behavior of that element, i.e. the constancy of
the Li/H ratio with metallicity, strongly suggests a primordial
origin. Its major isotope, $^7$Li, is indeed found to be produced in standard
primordial nucleosynthesis calculations, at a level close to
the one observed in halo stars (e.g. Lambert 2004 and references therein).

The other isotope of Li, \ls, is produced in extreme low levels in Big Bang
nucleosynthesis (\ls/H$<$10$^{-14}$, e.g. Serpico et al. 2004). Its only known 
source at present is non-thermal nucleosynthesis through the interaction
of CRs with the ISM. Its abundance is expected to rise continuously
during galactic evolution, similar to the one displayed by Be,
another light element solely synthesized by CRs. Taking into account 
the observed abundance of Be in stars of metallicity [Fe/H]$\sim$--3
and the respective production cross-sections, one expects that the
\ls/H ratio at such low metallicities would be considerably less than
10$^{-12}$.

The recent reports of the detection of  \ls  in halo stars
(Asplund et al. 2004, 2005) give a new twist to the 
LiBeB saga. The reported $^6$Li/H  value
at   [Fe/H]$\sim$--2.7 is much larger than expected if standard
galactic CRs are the only source of \ls. This problem was already 
noticed by Ramaty et al. (2000) after preliminary  
reports of \ls \ detection in very low metallicity halo stars. 
Equally suprising
is the report (Asplund et al. 2004, 2005) of a \ls \ plateau, at the
level of $^6$Li/H$\sim$10$^{-11}$ and
in the metallicity range --2.7 $<$ [Fe/H] $<$ --0.6; such
a plateau is reminiscent of the
Spite Li plateau and suggests a pre-galactic origin for \ls.
It should be stressed, at this point, that the reality of the $^6$Li plateau
or even its absolute level, is not well established at present. 
However, a recent preliminary analysis of {\it SUBARU} data 
(Inoue et al. 2005) corroborates the findings of Asplund et al. 
(2005) for $^6$Li in halo stars, although the  derived abundance 
could be twice as small.

Those intriguing, albeit not yet fully established, observational
results have already
prompted a few ideas using a pre-galactic origin
of that light isotope:

1) Primordial, non-standard production during Big Bang Nucleosynthesis
(Jedamzik 2004):
the decay/annihilation of some massive particle (e.g.
neutralino) releases energetic nucleons/photons that produce $^3$He or  $^3$H
by spallation/photodisintegration of  $^4$He, while
subsequent fusion reactions between $^4$He and $^3$He or  $^3$H 
create  $^6$Li. This scenario may have undesirable consequences 
on the abundances of other primordial nuclei (e.g. the D/$^3$He ratio),
as criticized by Ellis et al (2005), and will not be discussed here.

2) Pre-galactic by fusion reactions of  $^4$He nuclei, accelerated
by  the energy released during cosmic 
structure formation (Suzuki and Inoue 2002);
in that case, CR energetics are decoupled from the energetics of 
supernovae (SN), the latter bein at the origin of the failure of 
the conventional scenario (see Sect. 3.1). 

3) Finally, Rollinde et al (2004) postulated an early pre-galactic 
burst of CRs with
appropriately tuned  intensity in order to justify the reported level
of early \ls, but without considering potential sources or the
corresponding energetics.

All scenarii of \ls \ production involving energetic particles are constrained
by energy requirements: what is the source accelerating the energetic
particles (EP) and is the provided energy sufficient to justify the reported
abundance of \ls at very low metallicities? 
This is the main subject of this work. In Sec. 2 we
evaluate the energy requirements for \ls \ production. In Sec. 3 we 
compare these requirements with the energy potential of various candidate
sources of pre-galactic CR and  show that, in general, they are hard
to meet. Finally, in Sec. 4 we assume a pre-galactic \ls \ at the reported 
level and  study the evolution of the light element abundances
with  a detailed model of galactic chemical evolution, coupled to CR 
propagation and nucleosynthesis. In agreement with previous works, we find that
the existence of a \ls \ plateau is hard to justify, unless a fine-tuned 
metallicity-dependent depletion mechanism of \ls \ in stellar
envelopes is invoked. All those result together cast some doubt, either
on the reality of the reported \ls \ detection or on our understanding of
\ls \ production and evolution.

\section{Energy requirements for \ls \ production}

The energy requirements for the production of
light nuclei (Li, Be, B) through spallation of CNO nuclei
have been thoroughly studied in Ramaty et al. (1997).
Here a slightly different formulation of the problem is presented,
based on the formalism of CR propagation developed in MAR.

After acceleration  cosmic ray nuclei obtain an {\it injection
spectrum} $Q(E)$ as a function of energy $E$. 
While propagating through the ISM, they suffer
various losses (ionization, nuclear reactions, and escape from the Galaxy,
in the framework of the {\it leaky box} model).
The {\it propagated spectrum} $N(E)$ is assumed to reach 
{\it equilibrium} ($\vartheta N/\vartheta t$=0) rapidly:

\begin{equation}
\frac{\vartheta N(E)}{\vartheta t} \ = \ Q(E) \ - \frac{\vartheta}{\vartheta E} [ \ b(E) \ N(E) \ ] \ -
\frac{1}{\tau(E)} N(E) \ = \ 0
\end{equation}
where $b(E)$
represents ionization losses, and $\tau(E)$ is the effective timescale for
losses through nuclear reactions and escape from the Galaxy:
\begin{equation}
\frac{1}{\tau} \ = \ \frac{1}{\tau_{NUC}} \ + \frac{1}{\tau_{ESC}} 
\ \sim \frac{1}{\tau_{ESC}}.
\end{equation}

The functions $b(E)$ and $\tau(E)$ are determined from basic physics
and from the observed 
properties of the ISM (density, composition, ionization stage)
and of the CR (abundance ratios of primary to secondary and of unstable
to stable nuclei).
For {\it primary} nuclei, like H, He, C, N, O 
(the abundances of which are little affected by their propagation
through the interstellar medium), the solution of Eq.  1 is:
\begin{equation}
N(E) \ = \ \frac{1}{b(E)} \ \int_E^{\infty} Q(E') \ {\rm exp} 
\left[-\frac{R(E')-R(E)}{\Lambda} \right] \ dE' ,
\end{equation}
where 
\begin{equation}
R(E)=\int_0^E \frac{\rho \ v(E')}{b(E')} \  dE'
\end{equation}
is the {\it ionization range}, with $\rho$ the ISM mass density
and $v(E)$ the particle velocity, while
$\Lambda = \rho v \tau_{ESC}$ is the {\it escape length} from the Galaxy. 

The resulting equilibrium spectrum can also be expressed in terms of
the omnidirectional particle {\it flux} $\Phi(E) = N(E) v(E)$. 
That quantity is then folded with the relevant
spallation cross-sections $\sigma(E)$ in order to calculate the yield
of the LiBeB nuclei:
\begin{equation}
\frac{\vartheta y_k}{\vartheta t} \ = \sum_j y_j^{ISM} \ \sum_i \int_T^{\infty} \Phi_i^{GCR}(E)
\ \sigma_{ij}^k(E) \ P_{ij}^k(E_C) \ dE.
\end{equation}
In this expression, $y_k$ is the abundance (by number) of the light nucleus $k$
($k$=1,...,5 for $^6$Li,$^7$Li, $^9$Be, $^{10}$B,$^{11}$B). 
The indices $i$ and $j$ run over the range 1,...,5 for H, $^4$He, $^{12}$C, 
$^{14}$N, and $^{16}$O. The cross sections
$\sigma_{ij}^k(E)$ represent the probability of producing nucleus $k$
through the interaction of nuclei $i$ and $j$, and they have a threshold $T$.
The quantities
 $ P_{ij}^k(E_C)$ represent the fraction of light nuclei $k$ that are produced
at energy $E_C$ and are incorporated
in the ISM at time $t$. They are given by  
\begin{equation}
 P_{ij}^k(E_C) \ = \ {\rm exp}\left[-\frac{R_k(E_C)}{\Lambda}\right] 
\end{equation}
where $R_k(E)$ is the ionization range of nucleus $k$. The energy 
$E_C$ is close to zero when a fast proton or alpha hits a CNO nucleus
of the ISM (i.e. the resulting light nucleus is created at rest and $P\sim$1),
and $E_C=E$ when fast CNO nuclei are spallated by ISM protons and alphas
(i.e. the resulting light nuclei inherit the same energy per nucleon).
In the case of the fusion reaction $\alpha + \alpha \longrightarrow$ $^{6,7}$Li
($i=j=$2) the resulting Li nuclei are created with a velocity about half the
one of the fast $\alpha$ particles, and $E_C=E/4$ (see Eq. (6) in MAR).

The total power (energy per unit time) in accelerated particles is
\begin{equation}
\dot{W} \ = \frac{\vartheta W}{\vartheta t}  \ 
= \ \sum_i A_i \ \int^{\infty}_0 \ E \ Q_i \ dE,
\end{equation}
where multiplication by the mass number $A_i$ accounts for the
fact that energy $E$ is always expressed in units
of energy/nucleon. Obviously, by dividing Eq. (7) by Eq. (5) one obtains the energy of
accelerated particles of a given composition that is required to
produce one nucleus of species $k$. The result essentially depends
on two factors: the form of the injection spectrum $Q(E)$
and the  composition of that spectrum. The composition of the ISM is also 
involved, but it is always taken as equal to solar today, while
its evolution is constrained well by observations of low-metallicity stars.

\begin{figure}
\centering
\includegraphics[angle=-90,width=0.5\textwidth]{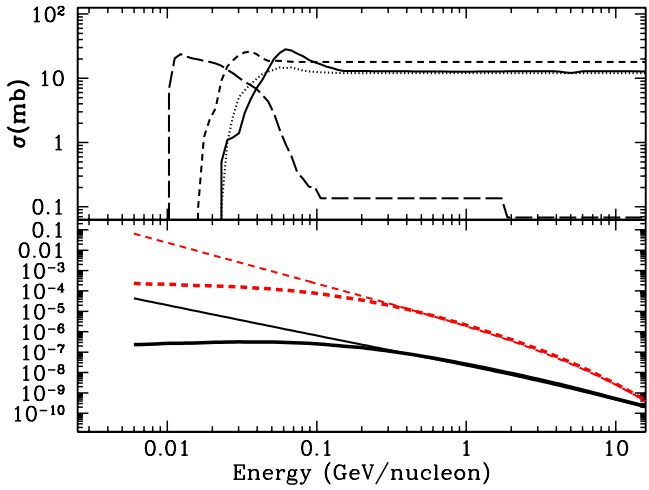}
\caption{{\it Top:} Cross sections for the production of $^6$Li from
proton spallation of $^{12}$C ({\it dotted}), $^{14}$N ({\it short dashed}),
$^{16}$O ({\it solid}) and $\alpha + \alpha$ ({\it long dashed}); the
recent measurements of Mercer et al. (2001) have been taken into account in
the last curve. {\it Bottom}: Energetic particle spectra adopted in this study;
they correspond to Eqs. (8) ({\it solid}) and (9) with $s$=3 and $E_0$=10 
GeV/nucleon ({\it dashed}).  In each case, the upper ({\it thin}) 
curve corresponds to the injection spectrum and the lower ({\it thick}) 
one to the equilibrium spectrum, propagated
in a leaky box model with an escape length $\Lambda$=10 gr cm$^{-2}$.
Normalization of the spectra is arbitrary.
} 
\end{figure}

The CR equilibrium spectrum is known very poorly at low energies, precisely 
those  that are important for Li production (in view of the relevant production
cross sections, see Fig. 1, upper panel). The reason is the poorly understood
modulation effects of the solar wind. Instead of using a demodulated spectrum
(e.g. Ip and Axford 1985), in most studies of Li production, a theoretical
injection spectrum is adopted and propagated in the Galaxy, in order to recover
the equilibrium spectrum through Eq. (3). The form of the injection spectrum is
motivated by theories of collisionless shock acceleration 
(e.g. Ellison and Ramaty 1985). 
Two popular spectra adopted in most studies in the field (Prantzos et al. 
1993, Fields et al. 1994, Ramaty et al. 1997, 2000) are

\begin{equation}
Q(E) \ \propto \ \frac{E+E_p}{[E(E+2E_p)]^{1.5}}
\end{equation}
where $E_p$=938 MeV is the proton rest mass-energy, and 

\begin{equation}
Q(E) \ \propto \ \frac{p^{-s}}{\beta} {\rm exp}(-E/E_0)
\end{equation}
where $\beta=v/c$ is the velocity expressed as a 
fraction of the light velocity,
$p$ the particle momentum per nucleon, 
the factor $s$ is usually 2$<s<$3 (in the case
of strong shocks), and $E_0$ is a cut-off energy. In view of the form
of the $\alpha+\alpha$ 
cross sections, one might think that
a much steeper spectrum than those two may favor the 
energetics of $^6$Li production. However, ionization losses increase
as $E^{-1}$ and are so
important in the energy range of few tens of MeV/nucleon that
steeper spectra lead to much larger energy demands. 
Even if the energy spectra
of energetic particles in the early Galaxy or in the pre-galactic era are
poorly known (see e.g. Gabici and Blasi 2003, Inoue et al. 2004), 
we feel that those adopted here represent 
the various possibilities reasonably well, at least 
as far as energetics is concerned (see below).

The adopted spectra appear in Fig. 1 (lower panel). The energetics of
$^6$Li production also depends on the adopted CR source 
composition. Today, that composition is very close to solar, once
effects of propagation and various biases are taken into account
(e.g. Wiedenbeck et al. 2001). Intuitively, it appears that the CR
source composition should always follow the one of the intestellar medium.
However, the observed linearity of Be vs Fe in Galactic stars (e.g. Primas et
al. 1999) strongly suggests that the CR composition varied little during
the Galactic history; this was first suggested by Duncan et al. (1992) and
convincingly demonstrated  by
 Ramaty et al. (1997) on the basis of energetics arguments.
Note that the composition of CRs in the early Galaxy affects 
the $^6$Li energetics  relatively
little, since a large fraction of that isotope
is produced by $\alpha+\alpha$ fusion reactions, especially at low 
metallicities (Steigman and Walker 1992).

\begin{figure}
\centering
\includegraphics[angle=-90,width=0.5\textwidth]{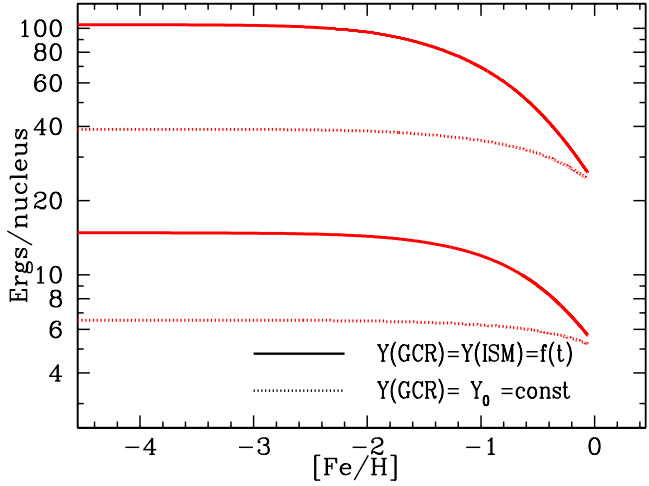}
\caption{Energy requirement for the production of $^6$Li, as a function
of the metallicity of the ISM, expressed by [Fe/H]. 
 The two upper curves correspond to
the spectrum of Eq. (8) and the two lower ones to the spectrum of Eq. (9)
with $s$=3 and $E_O$=10 GeV/nucleon (see Fig. 1).
The curves correspond to two different
assumptions about the composition of the energetic particles (EP) that hit
the ISM. The {\it solid curve} is obtained by assuming that the
composition of EP always follows the one of the ISM, i.e. it varies with
time. The {\it dotted curve} is obtained by assuming that the composition of
EP is always the same and similar to the source composition
of galactic CRs today. In the former
case,  $^6$Li at low metallicities is produced almost  exclusively by 
$\alpha + \alpha$ reactions. 
} 
\end{figure}

The discussion of this section is summarized in Fig. 2, where the energy
required to produce a $^6$Li nucleus by EP hitting the
ISM is displayed as a function of the evolving composition of the ISM,
represented by [Fe/H]. It is assumed that the abundances of
N and C in the ISM follow exactly the one of Fe, while the abundance of O
evolves differently 
(O/Fe) is 3 times solar for [Fe/H]$<$-1 and declines smoothly
to its solar value for higher [Fe/H]); this assumption is based on the
observed evolution of those elements.
The momentum spectrum of Eq. (9) is more efficient
than the energy spectrum of Eq. (8). Note that 
increasing the slope $s$ of the momentum spectrum to higher values
than the adopted value of $s$=3 
does not improve the energetics, because of the resulting high ionization 
losses\footnote{See e.g. Fig. 8c in Ramaty et al.
1997, displaying clearly the effects of slope $s$ and cut-off energy
$E_0$ on the energetics of Be; increasing $s$ (steeper spectra)
increases dramatically the energetic requirements. A similar, albeit not
so dramatic effect (in view of the large low energy $\alpha + \alpha$ 
cross sections)
holds for \ls.}. Thus, the two lowest curves
in Fig. 2 are close to the maximum efficiency for $^6$Li production. 
Pre-galactic CRs were obviously devoid of CNO nuclei, so the energy 
requirement for $^6$Li production in the case of pure $\alpha+\alpha$ is 
$\varepsilon_6\sim$16 erg/nucleus. We shall adopt it as a canonical value 
in the
following.[{\it Note:} Increasing the value of the escape 
length $\Lambda_{ESC}$ from 10 gr cm$^{-2}$ adopted here
to 100-200 gr cm$^{-2}$ (i.e. assuming a more efficient confinement of 
the CRs, which could be the case in the early Galaxy) slightly improves 
the energetics of $^6$Li production, by less 
than $\sim$40\%; however, this small 
difference does not affect the conclusions of Sec. 3].

The tentative plateau value in the data of Asplund et al. (2005)
is at ($^6$Li/H)$_{PL} \sim$10$^{-11}$ (where $PL$ stands for "`plateau"'). 
Taking the
discussion on energetics into account, one sees that $\varepsilon_6 \sim$16 
ergs in EP are required in order
to produce 1 nucleus of $^6$Li for every 10$^{11}$ nuclei of the ISM;
this energetic requirement can be expressed as
\begin{equation}
w_6 \ = \ \varepsilon_6 \ \left(\frac{\rm ^6Li}{\rm H}\right)_{PL}
 \ \frac{1}{m_P} \ \sim 10^{14} \  {\rm erg \ gr^{-1}}
\end{equation} 
where $m_P$ is the proton mass. This means that 
$\sim 10^{14}$ ergs of EP are required to pollute each gr of the ISM to the
level of  ($^6$Li/H)$_{PL} \sim$10$^{-11}$ through $\alpha+\alpha$ reactions.
We note that similar values are found
in Reeves (2005), through a different evaluation. 
In the next section we explore  the
energetic potential of various CR sources and compare it to the
requirement for early $^6$Li production of Eq. (10).
One should note, however, that if the true pregalactic \ls \ value
is higher than 10$^{-11}$ (see Sect. 4.2), the value of $w_6$ should be 
revised upwards accordingly.

\section{Candidate sources for particle acceleration in the
early Galaxy}

\subsection{Typical core collapse supernovae}

Core collapse supernovae release a ``canonical'' energy of 
$E_{SN}\sim$1.5 10$^{51}$ erg,
of which $\sim$10-20\% may accelerate CR particles, i.e. the efficiency
of turning SN shock energy into EP is $f_{SN}\sim$0.1-0.2. The efficiency value
results from the total power of Galactic 
CR, estimated to 2 10$^{41}$ erg s$^{-1}$ (e.g. Longair 1992), 
and from the statistics of supernovae in the Milky Way (about 3-4 supernovae
per century are expected on average, on the basis of observed SN frequencies
in Milky Way type galaxies, e.g. Mannucci  et al. 2005).

Taking  Eq. (10) into account, one sees that a mass of the ISM 
$M_{ISM} = \ f_{SN} \ E_{SN}/w_6 \sim$
10$^3$ \ms \ can be polluted by a single SN to the level of 
 ($^6$Li/H)$_{PL} \sim$10$^{-11}$. However, a ``canonical'' core-collapse SN,
 i.e. resulting from the explosion of a not too massive star, also
ejects $M_{Fe}\sim$0.07 \ms \ of Fe, (e.g. Woosley and Weaver 1995)
polluting the ISM to the level of 

$$ 
{\rm X_{Fe}} \ = \ \frac{M_{Fe}}{M_{ISM}} \ \sim 
$$  
\begin{equation}
5 \ 10^{-5} \frac{M_{Fe}}{0.07 \ M_{\odot}}
\left(\frac{f_{SN}}{0.1}\right)^{-1} \frac{w_6}{10^{14} 
{\rm erg/g}} \left(\frac{E_{SN}}{1.5 \ 10^{51} {\rm erg}}\right)^{-1}
\end{equation}
where ${\rm X_{Fe}}$ is the mass fraction of Fe.
This corresponds to a metallicity
[Fe/H]$\sim$-1.4 (adopting a solar mass fraction of 
${\rm X_{Fe,\odot}}$=1.25 10$^{-3}$ for Fe,
following Lodders 2003). Thus, ``canonical'' SN may indeed pollute the ISM
to the level of  $^6$Li/H$\sim$10$^{-11}$, but only for metallicities as
high as [Fe/H]$\sim$-1.4. At metallicities [Fe/H]$\sim$-2.7 
(the lowest metallicity
point in the data of Asplund et al. 2005), a simple scaling shows that the
expected level of pollution is only  $^6$Li/H$\sim$5 10$^{-13}$, i.e. a factor
of $\sim$20 below the observations. Note that this a real upper limit to the
level of $^6$Li enrichment that can be obtained by a "`canonical
"'supernova, because
it is assumed that  the EP accelerated by
 the supernova produce 
$^6$Li only inside the matter that is enriched in Fe. 
In actual reality, those EP diffuse much further and produce 
$^6$Li in a much larger region, but to a proportionally lower level (since 
the total number of $^6$Li atoms is determined by the energy of the explosion 
and is a constant: $\sim M_{ISM} (^6Li/H)_{PL}/ m_P \sim$2 10$^{49}$
 $^6$Li atoms).

The situation may be even worse
if the level of primordial $^7$Li is as high as suggested by the WMAP data.
Indeed, the baryonic density of the universe derived by observations of the
cosmic microwave background corresponds to 
$^7$Li/H$\sim$4 10$^{-10}$, according to standard calculations of 
primordial nucleosynthesis (e.g. Serpico et al. 2004). That value 
is 2-3 times higher than 
the observed plateau of Li/H in the low metallicity stars of the Milky Way, 
and the discrepancy may be attributed to our presently poor understanding 
of Li depletion in stellar envelopes (e.g. Lambert 2004 
and references therein). 
If that explanation is correct, an even greater depletion of the more 
fragile $^6$Li is expected; the true value of $^6$Li/H should then be 
at least 2-3 times higher than the one measured by Asplund et al. (2005), 
making the problem of its production by supernovae even worse.

\subsection{Energetic (and/or low Fe yield)  supernovae}

The  arguments of the previous section may not hold if the energetics of 
SN is decoupled from the Fe yield: if ($E_{SN}/M_{Fe}) >>$ (1.5 10$^{51}$ 
erg/0.07 \ms), Eq. (11) shows that a high $^6$Li/H could be obtained for 
a small [Fe/H]. Such a decoupling might be justified 
on both observational and theoretical grounds; however, as is usually the 
case, observations leave less room for optimism than does the theory. 

Observations
of extragalactic core collapse  SN suggest a clear, but not exactly linear,
correlation between  $M_{Fe}$ and $E_{SN}$, as shown in the review by
Hamuy( 2003). In Fig. 15 of that review, it can be seen that SN98bw, a Type Ic
supernova, may have ejected  0.5 \ms \ of Fe (i.e. 7 times the
canonical value adopted here) for an energy of 2-5 10$^{52}$ erg,\footnote
 {Iwamoto et al. 
(1998) derive an energy of 2-5 10$^{52}$ erg for SN1998b, while in Maeda and 
Nomoto (2003) the derived value is 3 10$^{52}$ erg.} which is 15-30 times the
 canonical energy of core collapse supernovae\footnote{
The large energy value of SN1998bw is derived by Iwamoto et al. (1998) under 
the assumption of spherical symmetry; however, it could be as low as only 
3 10$^{51}$ erg, if that assumption is dropped and an ellipsoidal geometry
for the ejecta is assumed, as suggested in  Hoeflich et al. (1998)}.
 If such events dominated in the early Galaxy, the discrepancy with
the energetics of \ls \ production would be reduced to $\sim$4-9 
(the original discrepancy of a factor of 20 is reduced by a factor of 15-30
because of the increased energy, and again augmented by a factor of 7
due to the increased Fe yield).
In their analysis, Nomoto et al. (2000) find that the progenitor mass 
of SN1998bw had $\sim$40 \ms. One might think that such energetic supernovae
with large $E_{SN}$/$M_{Fe}$ values would be common among massive stars
of the so-called "hypernova" branch (e.g. Nomoto et al. 2005). 
However, although
the energy of such explosions is indeed larger than the canonical value, the
$E_{SN}$/$M_{Fe}$ value is not always large. Thus, in the case of SN1999as, 
again a type Ic supernova, the energy is estimated as 3 10$^{52}$ erg 
(see e.g. Fig. 1 in Maeda and Nomoto 2003), but the Fe yield is also quite 
large, around 4 \ms, so that the resulting $E_{SN}$/$M_{Fe}$ ratio is even 
lower than in the  canonical case. Thus observations of energetic supernova
 in the nearby universe do not support
$E_{SN}$/$M_{Fe}$ ratios large enough to explain the $^6$Li/H plateau.

From the theoretical point of view, the relation between $E_{SN}$ and $M_{Fe}$ 
is impossible to derive at present, due to our very poor understanding of the 
explosion mechanism of core collapse supernovae (see e.g. 
Janka et al. 2003 for a review). 
In current 1D models of nucleosynthesis in supernovae, 
some assumptions have to be 
made about the way the shock wave induces explosive nucleosynthesis in the Si
layers (see e.g. Limongi and Chieffi 2003); comparison with observations 
(e.g. SN1987A in the LMC, a 20 \ms \ star that ejected 0.07 \ms \ of Fe) 
may then help to fix some of the parameters of the model, e.g. the so-called 
'mass-cut', the fiducial surface separating expanding material from material 
falling back to the compact object. Standard 1D calculations of that type 
usually produce a few 10$^{-2}$-10$^{-1}$ \ms \ of Fe for a few 
10$^{50}$-10$^{51}$ erg of kinetic energy, at least for stars in 
the mass range 12-35 \ms \  (e.g. Thielemann et al. 1996, Woosley 
and Weaver 1996,  Limongi and Chieffi 2003).

For larger stellar masses, in the range 30-100 \ms, simulations suggest that 
collapse to a black hole should be the general outcome, either directly or 
after the formation of a weak shock (see Heger et al. 2003 for a 
review of the situation for 1D simultions with no rotation). Even if 
the shock is weak, the almost total absence of metals in the ejecta of such 
explosions makes them suitable candidates for the production of 
early $^6$Li, free of the problem of the associated production 
of Fe. Further support for that possibility comes from the very short 
lifetimes of such objects (less than a few million yrs) and their 
potentially large number in a zero-metallicity stellar generation: 
indeed, some theoretical arguments suggest that the 
stellar IMF at zero metallicity was considerably 
skewed towards massive stars, in the 100 
\ms \ range (Nakamura and Umemura 2002). If such objects are at the origin of
the observed early $^6$Li/H, one can easily 
evaluate their population in the Milky Way halo (with 
a rather large uncertainty, though).

As discussed in Sec. 3.1, the mass of the ISM that a normal SN 
may ``pollute'' to the level 
of the observed early $^6$Li/H is $M_{ISM}\sim$10$^3$ \ms. We assume here
that  each of these explosions leaves a black hole with  mass $M_B \sim$50-100 
\ms. The stellar mass of the Galactic halo is $M_H\sim$2 10$^9$ \ms \ (Bullock
and Johnston 2005 and references therein),
and this may well be (within an order of magnitude) the mass of the
gas from which the halo was formed (see Fig. 1, right top panel, 
in Prantzos 2003). The number of black holes is then 
$N_B = M_H/M_{ISM}\sim$10$^6$,
and their total mass $N_B \times M_B \sim$10$^8$ \ms,
i.e. $\sim$10\% of the stellar  halo mass.
One immediately sees that only a ``top-heavy'' IMF can produce such a large
fraction of massive stars, 
since in a normal IMF (i.e. Salpeter or Kroupa et al. 1993), 
only $\sim$1\% of the mass is in 50-100 \ms \ stars. Unfortunately, the
required total mass in black holes is far below the microlensing detection 
limits
of the MACHO or EROS2 experiments: EROS2 places an upper limit of $\sim$20\% 
on the
fraction of the Milky Way's {\it dark halo} which could be in the form of such
 objects
(Tisserand, PhD Thesis 2004 and private communication). For a 10$^{12}$ \ms \
 dark halo,
this corresponds to $\sim$2 10$^{11}$ \ms. Thus, the idea that early $^6$Li 
is produced
by mildly energetic stellar explosions of a first generation of
(non-rotating) massive stars that lead to black holes and produce too few 
metals
does not violate current constraints from microlensing experiments. 

At this point, it should be noted that the two most metal-poor stars 
of the Galaxy (as far as their Fe content is concerned) are at present
HE 1326-2326 (a subgiant or main sequence star) and HE 0107-5240 
(a red giant).
They both exhibit low Fe abundances, [Fe/H/H]$\sim$-5.2, but quite large
C/Fe and N/Fe ratios (around 10$^4$ times solar) and high ratios of
 Na/Fe, Mg/Fe
and, rather surprisingly, Sr/Fe (around 10-100 times solar), while other 
abundance
ratios X/Fe are compatible with solar values
(Frebel et al. 2005). These peculiar abundance patterns
may indeed be explained (at least qualitatively) by ``faint'' SN, assuming that
some mixing of the inner Fe peak nuclei with the exterior layers has 
occurred and
that most of the Fe peak elements fall back onto the black hole. This
 mixing scheme,
dictated on purely observational grounds, might correspond to asymmetric
 explosions
of rotating stars (e.g. Nomoto et al. 2005). One might think that such 
explosions also
fulfill  the requirements of ``large energy and/or low Fe yield'' 
explosions for early
$^6$Li production. However, no Li has been found in HE 1327-2326 and only
 an upper
limit of $\sim$0.5 times the Spite plateau value is given in Frebel
 et al. (2005);
in all probability, $^6$Li should also be absent from its atmosphere. 
Of course, one might argue that $^6$Li was
present during the star's formation but subsequently depleted in its
envelope. In any case, there is no observational proof at present
that the supernovae responsible for
the abundance pattern of the most metal poor Galactic stars were at
the origin of the  early $^6$Li observed.

\subsection{Shocks from cosmic structure formation}

The currently popular scenario of hierarchical structure formation in the 
Universe 
suggests that large scale objects, such as galaxies and clusters, are formed
 from 
the merging of smaller subsystems, which are moving and virialised
in the gravitational potential wells of dark matter haloes.
 During those mergers, 
shocks should develop in the gaseous component of the merging subsystems. 
The kinetic energy of those shocks from structure formation should be, on 
average,
\begin{equation}
E_{SF} \ = \ \frac{1}{2} \ M \ v_{VIR}^2
\end{equation}
where $M$ is the mass of the baryonic gaseous component  and the virial 
velocity is
\begin{equation}
v_{VIR} \ \sim \ \left(\frac{G M_{DH}}{R}\right)^{1/2} \ \sim \ 400 
\ \left(\frac{M_{DH}}{10^{13} M_{\odot}}\right)^{1/4} \ {\rm km/s},
\end{equation}
$G$ being the gravitational constant, $M_{DH}$ the dark halo mass, 
and $R$ the virial radius. The numerical values in Eq. (12 ) result from
 cosmological simulations 
of large-scale structure formation (Christophe Pichon, 
private communication). In  the case 
of the Milky Way, the present-day dark halo mass is evaluated to $M_{DH} 
\sim$10$^{12}$ \ms (Battaglia et al. 2005), leading to a virial velocity
$v_{VIR, MW} \sim$220 km/s today.

The energy of those shocks {\it per unit mass of gas} is 
$w_{SF}$=1/2 $v_{VIR}^2$, 
and for the case of the Milky Way one gets 
$w_{SF,MW} \sim$2 10$^{14}$ erg gr$^{-1}$. In order to satisfy the energetics 
requirement for $^6$Li production of Eq. (10), the kinetic energy of shocks 
from the formation of the Milky Way should be converted to energetic particles 
with an efficiency of 50\%, which is an extremely high value. Suzuki and Inoue 
(2002), who first suggested the idea that shocks 
from cosmological
structure formation may be at the origin of early $^6$Li production,
assumed a higher value for  $M_{DH}$ 
(3 10$^{12}$ \ms) and found a reasonable acceleration efficiency
 of 15\%. However,
this estimate is overly 
optimistic, because it is based on the assumption that  {\it  the dark halo is 
fully formed before the first stars appear polluted by } $^6$Li. Taking 
 the age of the low metallicity halo stars of the Milky Way into 
account (larger  than 
11 Gyr, corresponding to a redshift $z>$3), such a hypothesis is extremely 
improbable. 
Indeed, a simple application of the  Press-Shechter formalism (e.g. in the 
Appendix
of van den Bosch 2002) 
shows that a dark halo of 10$^{12}$ \ms \ is only assembled at redshift 
$\sim$1.1-1.3. 
In the earliest phases of our  Galaxy, only much smaller dark haloes 
(less than 10$^{10}$ \ms) may have existed; the corresponding virial 
velocity being lower than 200 km/s, the resulting kinetic energy was 
too low to fulfill the energy requirement for $^6$Li production.
 
One may, in fact, obtain an upper limit to that energy by noting that
the mass of the baryonic halo today is $M_H\sim$2 10$^9$ \ms.
Assuming
a baryonic/dark matter ratio of 0.1, this corresponds to a dark matter halo
of 2 10$^{10}$ \ms \footnote{A recent work (Bullock and Johnston 2005), 
that studies the formation of Milky Way's halo in a cosmological context
confirms that the halo was indeed formed  by the merging and tidal disruption 
of dwarf galaxies, mostly embedded in dark haloes of $\sim$10$^{10}$ \ms.} 
and to a corresponding virial velocity of $\sim$100 km/s
(from Eq. 13). This leads to a total kinetic energy per unit mass of 
baryons of
$\sim$0.5 10$^{14}$ erg gr$^{-1}$ and, adopting a standard conversion 
efficiency
to energetic particles ($f\sim$0.1), one sees that no more than 5\% of the
observationally required $w_6$ value can be provided by that mechanism.
Thus, shocks from early structure formation cannot be at the origin of
early $^6$Li.

Note that, in the framework of the hierarchical structure formation paradigm,
mergers occurred during a large fraction of the Galaxy's history (e.g. Helmi et
al. 2003). If the corresponding shocks accelerated EP, the proposed
scenario cannot be characterized as ``pre-galactic'' and it would
lead to a continuous rise of the $^6$Li/H ratio, not to the observed 
``plateau''. A metallicity-dependent mechanism of $^6$Li depletion
inside stars should then also be introduced to account for the
``plateau'' (see also Sec. 4.2).

\subsection{Accretion onto the Galactic black hole}

Accretion onto black holes may have provided another energy source 
for particle acceleration
in the early galactic (or pre-galactic) era. Both theory and observations
suggest that a large fraction of the black hole mass-energy may be extracted
in that case, either in the form of a jet or a wind; this fraction may
be as large as $\eta\sim$0.1. 

The largest black hole in the vicinity of the Milky Way is the one laying 
in the Galactic center (GC). Its mass is estimated to $M_{BH}\sim$3 10$^6$ \ms \ 
(Melia and Falcke 2001). The energy that could be released by accretion
onto it is at most $E_{BH}$= $\eta M_{BH} c^2$, where $c$ is the light
 velocity.
Assuming a high value for the energy extraction efficiency ($\eta$=0.1)
and a typical value for the efficiency of conversion of that energy to
energetic particles ($f$=0.1) one finds  that
\begin{equation}
E_{BH} \ \sim \ 5 \ 10^{58} \ \frac{\eta}{0.1} \ \frac{f}{0.1} \ 
\frac{M_{BH}}{3 \ 10^6 
{\rm M}_{\odot}} \ {\rm erg}.
\end{equation}

Assuming that the black hole was already in place before the
formation of the Milky Way's halo, 
one finds that a mass of $E_{BH}/w_6 \sim$2 10$^{11}$ \ms \ could have been
polluted to the level of the observed early $^6$Li/H. This is $\sim$3 times
larger
than the total stellar mass of the Milky Way (around 6-7 10$^{10}$ \ms,
including the bulge), comparable to the total stellar mass
of the Local Group, which is dominated by the Milky Way and Andromeda,
and 100 times larger than the stellar mass of the 
Galactic halo
$M_H\sim$2 10$^9$ \ms. Thus, the Galactic center black hole apparently fulfills
the energy requirement of Eq. (14) for early $^6$Li production; 
however, this
occurs only under extreme assumptions about 1) the time of its formation and
2) the efficiency of extracting its mass-energy.

The first of those points is, perhaps, not that crucial: observations
of luminous quasars at high redshift (e.g. in the Sloan Digital Sky Survey,
Fan et al. 2004) imply that massive black holes
were already in place in the first billion years of the cosmic evolution.
Similar conclusions are drawn in a recent study of the accretion history
of supermassive black holes based on observations of the current population
and accretion rates of those objects (Hopkins et al. 2005b).

The second assumption is perhaps more difficult to justify, in view of the
well-known (and poorly undestood) inefficiency of the Galactic
black hole for converting accreted matter into radiation. Indeed, the bolometric
luminosity of Sgr A$^*$ is $\sim$10$^5$ times lower than what is expected 
from the well-determined Bondi accretion rate of $\sim$10$^{-5}$ \ms/yr
(e.g. Feng, Quataert and Narayan 2003 and references therein). If the same
low efficiency characterised the GC black hole in its earliest life (i.e.
if $\eta$=10$^{-5}$ in Eq. 14), and if we assume
that the kinetic energy
of the matter escaping the black hole (in the form of a jet or wind) is 
equivalent to the energy radiated by the black hole, then
the corresponding energetic particles
could enrich  just the mass of a large globular cluster
($\sim$10$^6$ \ms)  with $^6$Li. Of course, one may assume that, for some
reason, the efficiency
of the GC black hole was much higher in its early youth, in which case 
the energetics of early $^6$Li production could be satisfied.
However, even in that case, it is difficult to imagine that cosmic rays
accelerated from a single object could produce a uniform abundance
of \ls \ over the whole Galactic volume (a radial gradient is rather expected).

The conclusion of this section is that, although promising in principle,
the idea that accretion on the GC black hole is at the origin of early
$^6$Li encounters some serious difficulties.

\subsection {Accretion onto supermassive black holes in the Universe}

On a larger scale than the one of the Milky Way, accretion onto 
supermassive black holes may provide an important source of energy
for particle acceleration in the Universe. Such objects are now routinely 
found in the centers of galaxies (see Ferrarese and Ford 2005 for a 
recent review). One may obtain a useful upper limit to their capacity to
produce significant amounts of \ls \ by considering the following:

The present day cosmic density of those objects (in units of the critical
density) is evaluated to $\Omega_{SMBH}$=4 10$^{-6}$, while the corresponding
total baryon density is $\Omega_{BAR}$=4.5 10$^{-2}$ (Fukugita and Peebles
2004). As in the previous subsection, assuming that a fraction $\eta\sim$0.1 
of the black hole rest mass can be usefully extracted as mechanical/radiative
energy, and that a fraction $f\sim$0.1 of it can be used in accelerating
particles, one sees that the corresponding energy in such particles per
unit baryon mass is:
$$
w_{SMBH} \ = \frac{\eta \ f \ \Omega_{SMBH} \ c^2}{\Omega_{BAR}} \ =
$$
\begin{equation}
10^{15} \  \frac{\eta}{0.1} \ \frac{f}{0.1} \ 
\frac{\Omega_{SMBH}/\Omega_{BAR}}{10^{-4}} \ {\rm erg \ g^{-1}}
\end{equation}
i.e. $w_{SMBH} \sim$10 $w_6$.
It then appears that supermassive black holes should be able to provide the
energy needed to produce early 
($^6$Li/H)$_{PL}$, since the available energy per unit baryon mass 
is about ten times larger than
required. However, one should again consider the meaning of those numbers.
They imply that those supermassive black holes were fully formed 
and released the accreted energy  {\it before the formation of the first stars
of the Galactic halo}, i.e. at redshifts $z>3$.

The recent study of Hopkins et al. (2005b), based on observations of the nearby
supermassive black hole population, indeed suggests  that the bulk
of the supermassive black hole mass was accreted early on in a radiatively
efficient accretion phase. However, it is not clear exactly when that happened.
For instance, studies of the evolution of the quasar population suggest that
the maximum in their number density in any luminosity interval was at 
redshift $z<$2 (see e.g. Fig. 4 in Hopkins et al. 2005a). 
If this evolution also characterizes  the formation history and accretion rate
of supermassive black holes, then those sources cannot be at the origin
of observed early \ls.

\section{Evolution of \ls }

In this section we {\it assume} that $^6$Li was already present in the earliest
moments of the formation of the Milky Way, at an abundance level 
at least as high as suggested by
the observations of Asplund et al. (2005). We study its subsequent evolution
with a detailed model of galactic chemical evolution, including
 its production by fusion of energetic alpha particles
and spallation of CNO nuclei (see Sec. 4.2) 
in  a self-consistent way; our aim is not to fully reassess
the whole subject of LiBeB production (which is still poorly
understood, despite the large amount of theoretical work devoted to it;
see Prantzos 2004 for a short review) but rather to study the
implications of a pregalactic $^6$Li component.

\subsection{Pre-galactic $^6$Li: cosmic or just local ?}

At this point, a (quite useful)  distinction should be made between the terms
primordial, cosmic pre-galactic, and local pre-galactic.

The term {\it primordial} implies production of $^6$Li in the early Universe,
either during the period of Big Bang nucleosynthesis or shortly after,
i.e. through the decay of an unstable (super-)particle e.g.
Jedamzik (2004). The resulting abundance of $^6$Li is then characteristic of the
{\it total baryonic content} of the Universe, i.e. of the baryonic fraction
$\Omega_{BAR}$. 

The term {\it cosmic pre-galactic} implies production of $^6$Li
prior to star or galaxy formation {\it everywhere} in the Universe; i.e. 
the resulting abundance of $^6$Li is again characteristic of the
{\it total baryonic content} of the Universe, so both the 
intergalactic medium (IGM) and the star-forming galaxies 
have the same $^6$Li/H ratio. This is assumed in e.g. 
the scenario of Rollinde et al. (2005). 

The term {\it local pre-galactic} implies that $^6$Li has only
(or mostly) polluted the baryonic gas that {participates in galaxy
formation} and not at all  the baryons of the
intergalactic medium (or very little). This picture corresponds to the realistic case
where accelerating sources and the energetic particles producing $^6$Li
are mostly confined
inside the high density gas that decouples from the Hubble flow and
forms galaxies. Indeed, except for the case of supermassive black holes,
all the energy sources explored in Sec. 3 
(including shocks from structure formation) belong
to this class. Moreover, the magnetic field is much more intense
within the dense gas surrounding such sources than in the rarefied IGM,
and it certainly produces some local confinement, albeit to a degree that
is very hard to evaluate at present (in view of our poor understanding
of the origin and evolution of galactic magnetic fields).

The distinction between {\it cosmic} and {\it local} pre-galactic is 
important for reasons related to the energetics of $^6$Li
production.
The local scenario requires less energy than the cosmic one, 
since in the former case only a fraction of the baryonic matter 
(the one participating in early galaxy formation) is involved. 
For instance, one can assume that only the 
$\sim$2 10$^9$ \ms \ of the halo mass was polluted with ($^6$Li/H)$_{PL}$ 
early on; the $\sim$5 10$^{10}$ \ms \ of the Milky Way disk were accreted 
much later from the intergalactic medium and  were enriched to much 
lower levels of $^6$Li (produced from the small fraction of energetic 
particles that escaped confinement in the halo region).

\subsection{Evolution of \ \ls \ and Be in the Milky Way}

The evolution of the light isotopes $^{6,7}$Li, $^9$Be, $^{10,11}$B is 
followed with a detailed model of the Milky Way's chemical evolution. 
The model, presented in Goswami and Prantzos (2000), satisfactorily reproduces 
 all the major observational constraints in the solar 
neighborhod, and in particular, the metallicity distributions of 
halo and local disk stars. The only difference with that model 
is that the  recent, metallicity dependent,  massive star yields 
of Chieffi an Limongi (2003) are adopted here. They differ from 
those of Woosley and Weaver (1995), which were adopted in GP2000, 
in several respects, and in particular in the absence of 
neutrino-induced nucleosynthesis, so no primary $^{11}$B and $^7$Li 
are present in the CL2003 yields. However, the yields of C, N, O, and Fe, 
most important for following the overall metallicity and the evolution 
of ligth isotopes by spallation, display only small differences
 between the two sets (see Goswami and Prantzos 2003 for a comparison).

\begin{figure}[!t]
\centering
\includegraphics[width=0.4\textwidth]{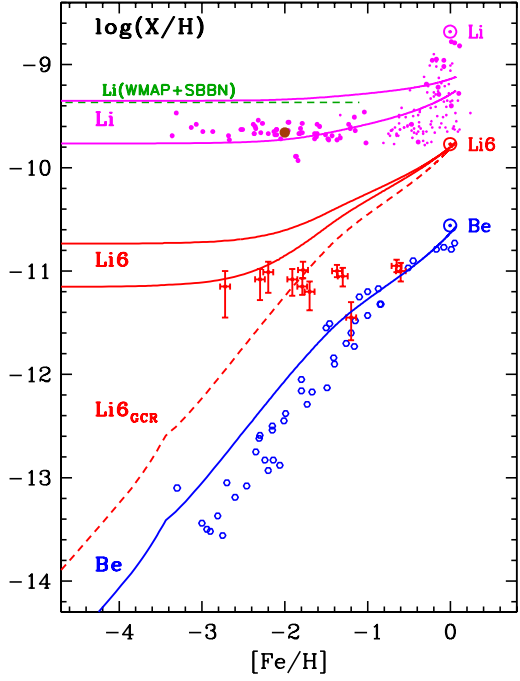}
\caption{Evolution of total Li, $^6$Li, and Be in the Milky Way.
Observations for Li  are: at
low metallicities from Melendez and Ramirez (2004, {\it small dots}) 
and from Bonifacio et al (2002, {\it large dot}); 
and at high metallicities from Chen  et al. (2001, {\it small dots}) 
and Boesgaard and King (1993, {\it large dots}). At high metallicities,
only values of Li/H$>$--9.8 are plotted, to avoid confusion with
$^6$Li values. Observations
for Be ({\it open symbols}) are from Garcia-Lopez et al. (1998)
and Primas et al. (1999). Observations for $^6$Li 
({\it with error bars}) are from Asplund et al. (2005)
and references therein. The Li abundance  corresponding
to the baryonic density of the Universe derived by WMAP is indicated
as {\it dashed horizontal line}. 
The   curves correspond to a simple chemical evolution model
with metallicity dependent stellar 
yields, GCR composition {\it assumed primary}
(to reproduce the Be observations)  and GCR spectra given by Eq. (9) with
$s$=3.  The contribution of the GCR component
of $^6$Li is indicated by a {\it dashed curve}.
The primordial $^7$Li is assumed to be either low (at the
level of the plateau of Charbonnel and Primas 2005, slightly lower than
derived by Melendez and Ramirez 2004) or high
(at the level of WMAP); only production
by GCR is considered for Li7, i.e. no stellar source. 
Similarly, the pre-galactic
$^6$Li is assumed to be either ``low'' (at the level observed
by Asplund et al. 2005), or ``high'' (by 0.$\sim$0.4 dex, for consistency with
assumed high Li value, which implies depletion of \ls 
in the stellar envelopes by at least that value)
} 
\end{figure}

The production of the light isotopes is followed as in Prantzos et al. (1993),
i.e. using the formalism presented in Sec. 2 and, in particular, the momentum 
injection spectrum of Eq. (9). A major difference with Prantzos et al. (1993) 
is the adopted normalization precedure for the light element abundances: 
instead of normalizing the model Be abundance 4.5 Gyr ago to its solar 
value (and scaling  all other abundances of light elements accordingly)  
we assume here 
that a fraction $f_{SN}$=0.1 of the kinetic energy $E_{KIN}$=1.5 10$^{51}$ 
erg of each SN is used to accelerate CR particles, i.e.
\begin{equation}
\dot{W} \ = \ f_{SN} \ E_{KIN} \ R_{SN}
\end{equation}
where $\dot{W}$ is given by Eq. (7) and $R_{SN}$ is the SN rate (number of 
SN per unit of time) given by the model. As shown by Ramaty et al. (1997), 
this procedure is the only one that guarantees consistency between the 
energetics of CRs and SN. Another difference with Prantzos et al. (1993) 
is that the light isotopes are {\it assumed} to be produced as primaries 
by CR, i.e. that the CR composition does not vary with time. This is, 
indeed, the only way to reproduce the observed linearity between Be and 
Fe (see Fig. 3), since secondary production fails energetically at low 
metallicities. Even the total SN energy turned into CR is not sufficient
 to produce the required number of Be atoms, as convincingly argued by 
Ramaty et al. (1997). It should be noted though that the reason for such 
 $\sim$constant CR composition has not been satisfactorily 
explained up to now (see Prantzos 2004 for a short critical assessment).

The results of our calculation concerning the evolution of Be, \ls, 
and total Li, appear in Fig. 3. We adopt a pre-galactic Li value that is 
either "low"', i.e. at the level  of the
low-metallicity Li plateau reported by Charbonnel and Primas (2005), or 
"high", i.e. at the level suggested by WMAP data plus Standard Big Bang 
Nucleosynthesis. In the latter case, currently observed Li in halo stars
 has been depleted by about 0.4 dex. Similarly, and for consistency, 
a "low" and a "high" value are adopted for  pregalactic
$^6$Li, respectively \ls/H=10$^{-11}$ (at the level of the "plateau"
reported by Asplund et al. 2005) and 0.4 dex higher (assuming its
 depletion has been equal to the one of total Li). Note that the latter 
value corrsponds to the minimal possible amount of \ls \ depletion,
since this isotope is more fragile than $^7$Li and should be more
depleted (see Table 6 in Asplund et al. 2005, based on calculations
by Richard et al. 2005). It can be seen that:

- The evolution of Be is satisfactorily reproduced; both its solar value and 
the slope of Be vs Fe are reproduced with the energy normalization of Eq. (15).
We stress, however, that the slope is ``naturally'' produced only under
the assumption of a time invariant CR composition (as suggested first
by Duncan et al. 1992), which has no sound theoretical justification at 
present.

- The CR component of \ls is sufficient to produce the solar value 
of that isotope (as already found analytically in MAR) but fails to reproduce
the lowest metallicity value reported by Asplund et al. (2005)
by  a factor of $\sim$10 (compared to the factor of 20,
which was analytically derived in Sec. 3.1).

- Assuming that the \ls \ plateau is real and extends to metallicities
as high as [Fe/H]=--0.6, one sees that the CR component of \ls \ alone (dotted
curve in Fig. 3)
crosses that plateau value slightly earlier (around [Fe/H]=--1.8).
A depletion mechanism {\it depending on metallicity} should be
introduced to justify a plateau in the range --1.8$<$[Fe/H]$<$ --0.6
(see also Rollinde et al. 2005).
When the assumed pre-galactic \ls \ component is also taken into account 
(either ``low'' or
``high''), the \ls \ abundance curve leaves the
plateau value even earlier, around [Fe/H]=--2.4. 

\begin{figure}[!t]
\centering
\includegraphics[width=0.4\textwidth]{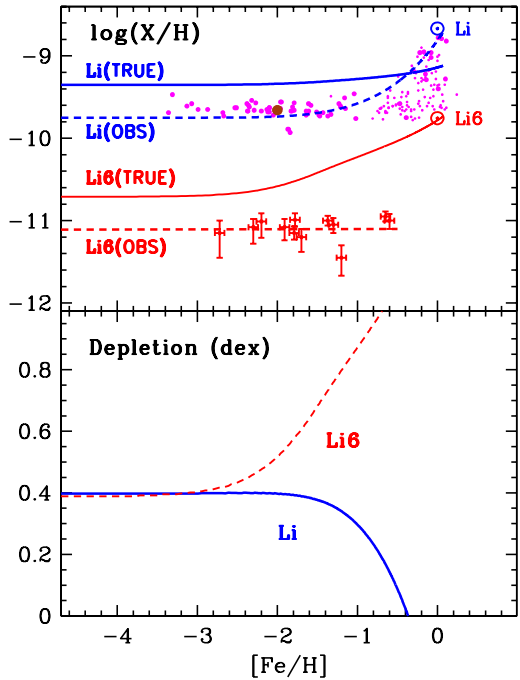}
\caption{{\it Upper panel:} Evolution of Li and \ls, observations vs 
theoretical estimates. Data points are as in Fig. 3. {\it Dashed}
curves corresponding
to observations are: for \ls, at the plateau level suggested by Asplund et al. 
(2004, 2005, and references therein), 
for metallicities lower than [Fe/H)=--0.6; and for Li, the
plateau value of Charbonnel and Primas (2005) for halo stars, 
plus the upper envelope
of the data points for higher metallicities (of course, there is no unique 
way to draw that envelope, especially because of the lack of data in the
region around [Fe/H]=--1). {\it Solid} 
curves corresponding to the ``true'' run of the
abundances (as expected from theoretical considerations) are the upper
curves of Fig. 3 for both Li and \ls. Note, however, that no
late stellar production is considered for Li; consequently, the 
corresponding curve underestimates the ``true'' Li abundance in disk
stars.
{\it Lower panel:} Corresponding depletion that 
Li and \ls \ have suffered in stellar 
envelopes. Depletion of \ls \ at the lowest 
metallicities is assumed to be equal to depletion of Li (i.e. the difference
of $\sim$0.35 dex
between values found by WMAP data analysis 
and by Charbonnel and Primas 2005), which is the
minimum possible value for \ls depletion 
in view of the greater fragility of \ls.
}
\end{figure}

A mechanism that depletes \ls \ in the stellar envelope with a metallicity
dependent efficiency is, perhaps, not all that hard to find. Indeed, it
is well known that convection sets in more easily in stellar envelopes with
higher opacity, that is, higher metallicity. More metallic stars have
deeper convective envelopes, which could bring \ls \ to higher temperatures
and destroy it more efficiently. However, it is much harder
to imagine that such a mechanism preserves  the putative \ls \
plateau value  {\it exactly}. Also, note that
depletion of total Li is mandatory when one assumes a high primordial 
Li abundance (as suggested by WMAP, see Fig. 3) 
in order to bring that value to the observed plateau level of Li; here
the depletion mechanism is metallicity independent for 
halo stars, but it also has
to maintain a plateau value, without significant dispersion (see 
Lambert 2004 and Charbonnel and Primas 2005).

Assuming that both Li and \ls \ have been equally 
depleted from their pre-galactic
values in the atmospheres of the lowest metallicity stars and 
that the observed plateau values are those given by Charbonnel
and Primas (2005)
for Li and by Asplund et al. (2005) for \ls, one can  quantitatively trace the
evolution of the stellar depletion of those elements, which is required
to keep their abundances within the observed levels. This is done
in Fig. 4. For total Li, the ``observed'' abundance level is situated 
at the plateau value for halo stars plus the upper envelope of the data for
disk stars (the sparsness of data around [Fe/H]=--1 making  a 
precise definition of the transition metallicity difficult). As can be seen in the lower 
panel of Fig. 4, the depletion of \ls has to be increasing with metallicity
in order to preserve the observed plateau, from the original depletion value
of $\sim$0.4 dex to almost its double at [Fe/H]=--1.2, and to about three times
a higher value at [Fe/H]=--0.6. On the other hand, depletion of Li has to remain 
essentially constant in halo stars, but necessarily to decrease with
metallicity at some point. This is even truer if it is confirmed that the
Li abundances in halo stars do not form a real plateau but increase slightly
with metallicity (e.g. Ryan et al. 1999, Boesgaard et al. 2005, Asplund et al. 2005,
Charbonnel and Primas 2005).

Stated  in a different way, Fig. 4 suggests that
standard CRs play an important early role
in the case of \ls, but not  in the case of the much more abundant $^7$Li;
the primordial component dominates the abundance of the latter in halo stars.
In order to cancel the effect of the CR contribution and to keep
\ls at the level of the  observed ``plateau'',
stellar depletion has to be progressively greater in the case
of \ls. In the case of Li, a metallicity independent
depletion (or slowly decreasing with metallicity) has to be invoked,
to bring agreement  the primordial (WMAP) value and observed
plateau values into agreement. Whether a realistic stellar environment can indeed
produce such a differential (and fine-tuned to preserve the plateau values)
depletion, remains to be discovered.

\section{Summary}

In this work, we reassess the problem of \ls \ evolution in the Milky Way,
motivated by the recently reported existence of a ``plateau''-like
behavior of its isotopic abundance  
in old and metal poor halo stars (Asplund et al. 2005).

At first, we calculate the energy requirements for \ls \ production
through fusion and spallation reactions, during the propagation of
energetic particles in the ISM.
We show that, even under most favorable conditions (i.e. steep 
CR spectra), it takes at least 10$^{14}$ erg gr$^{-1}$ 
to justify the reported value of (\ls/H)$_{PL}\sim$10$^{-11}$
(10$^{14}$ erg in accelerated particles for each gr of the ISM).

We proceed then by examining the energy performances of various
candidate acceleration sources that may have operated in the early
Galaxy: normal core collapse SN, atypical SN (energetic and/or having
a low Fe yield), shocks from cosmic structure formation (an interesting
suggestion of Suzuki and Inoue 2002), and the supermassive black hole
lying in the Galactic Center. We find that:

- normal SN producing $\sim$0.07 \ms \ of Fe could satisfy the energetics,
but they should also  simultaneously 
enrich the ISM to a level of [Fe/H]$\sim$--1.4;
they fail to produce the earliest reported \ls \ value (at  [Fe/H]$\sim$--2.7) 
by a factor of $\sim$10.

- energetic SN with low Fe yield could certainly satisfy all the 
\ls \ constraints; we note, however, that most energetic SN observed
today (sometimes called ``hypernovae'') have rather high Fe yields
and, despite their high energy, they fail for the same reason as normal SN.

- shocks from structure formation are not powerful enough, since in the
early times of Galaxy formation, the masses of the assembling dark haloes
were still quite small and the corresponding virial velocities insufficient;
in the most optimistic case (towards the end of halo formation), derived
energies are lower than required 
by a large factor (on the order of 20, i.e. they fail by a factor
larger than  normal SN).

- the Galactic black hole satisfies the energy requirements, assuming
that a) it is formed to a large extent even before the metal poor stars
of the halo (perhaps not an unreasonable assumption) and b) its efficiency
in converting accretion energy into accelerated particles was then
much higher than its present-day notorious inefficiency in turning 
accretion energy into radiation.

- supermassive black holes in galaxies could pollute the total baryonic content
of the Universe with ($^6$Li/H)$_{PL}$, provided that they
they were mostly formed before the formation of the very low-metallicity
halo stars. This requirement is not compatible with recent
ideas about the quasar luminosity evolution (Hopkins et al. 2005a), suggesting
that the peak in the number density of those objects occurred at redshift $z<$2.

We note that, in the case of early \ls,  a useful distinction should be made
between {\it cosmic} and {\it local} pre-galactic values.
The former implies pollution of the total baryonic content of the Universe,
where the latter only concerns the baryons assembled in galaxies; obviously, 
the energy requirements are easier to fulfill in the latter case than in
the former. 

Finally, we study the evolution of the light elements with a full-scale
galactic chemical evolution model, which satisfies all the major observational
constraints. We assume momentum CR spectra, time-independent CR composition,
a pre-galactic Li value either ``low'' 
(at the level reported by Charbonnel and Primas 2005) or ``high''
(WMAP value)  
and a pre-galactic \ls \ value again either ``low''
(at the level reported by Asplund et 
al. 2005) or ``high'' (by $\sim$0.4 dex, minimal amount of depletion to
be consistent with the assumption of ``high'' Li).

The model reproduces  the observed linearity between
Be and Fe abundances ``by construction''. 
The associated production of \ls \ by CR is
found to ``break'' the reported \ls \ plateau as early as [Fe/H]=--2.4.
A fine-tuned and metallicity-dependent mechanism of \ls \ depletion in stellar
envelopes would then be required in order to preserve the plateau value.
We  quantitatively evaluate the amount of required stellar depletion, and
show that in the case of \ls, it should increase with metallicity, while
for Li it should roughly be metallicity-independent (or even decreasing
with metallicity).

In summary, the present study suggests that 
1) the energy requirements for large early \ls \ production are
very constraining and hard to fulfill by the currently suggested sources, 
and that
2) contrary to the case of Li, the reported 
presence of a \ls \ plateau in halo stars
is  ``threatened'' by the  production of \ls \ by ordinary CRs (the same that produce
the observed Be) and requires some ``fine-tuning''. In view of these
implications, an unambiguous determination of the presence of \ls \ in halo
stars (absolute abundance values and reality of the plateau) 
is urgently required. In that respect, we note that Inoue et al. (2005) 
 recently report tentative detection of \ls \ in halo stars, at a level
roughly compatible with the one   reported by Asplund et al. (2005).

\acknowledgements{ I am grateful to H. Reeves, S. Inoue, and K. Jedamzik 
for several useful discussions.}

\def\aj{AJ }
\def\apj{ApJ }
\def\apjs{ApJS }
\def\aap{A\&A }
\def\aaps{A\&AS }


{}

\end{document}